\documentclass{amsart}  
\usepackage{amssymb}  
\usepackage{latexsym}  
  
\newcommand{\Z}{{\mathbb Z}}  
\newcommand{\R}{{\mathbb R}}  
\newcommand{\C}{{\mathbb C}}  
\newcommand{\N}{{\mathbb N}}

\newtheorem{theorem}{Theorem}     
     
\newtheorem{lemma}{Lemma}[section]     
\newtheorem{prop}[lemma]{Proposition}

\newcommand{\tr}{{\mathrm{tr}}}  
  
\newcounter{smalllist}

\begin{document}  
\title[Uniform dynamical bounds for the Fibonacci Hamiltonian]{Uniform dynamical bounds for the\\Fibonacci Hamiltonian}  
\author{David Damanik}  
\thanks{Research partially supported by NSF Grant DMS--0010101} 
\maketitle  
\vspace{0.3cm}      
\noindent      
Department of Mathematics 253--37, California Institute of Technology,      
Pasadena, CA 91125, U.S.A.\\[2mm]      
E-mail: \mbox{damanik@its.caltech.edu}\\[3mm]      
\begin{abstract}  
We prove quantum dynamical upper bounds for operators from the Fibonacci hull. These bounds hold for sufficiently large coupling and they are uniform in the phase. This extends recent work by Killip, Kiselev and Last who obtained these bounds for one particular phase. The main ingredient in our proof is a detailed combinatorial analysis of the sequences in the Fibonacci hull.
\end{abstract}  
  
\section{Introduction}

Quantum dynamics in the presence of purely singular continuous spectral measures has been the object of quite intense recent research activity. It has been shown that apart from classical results (usually referred to as the RAGE theorem), only little can be said about the associated dynamics in general and hence more detailed information on the spectral measures is required to obtain further restrictions on the dynamical behavior. 

An important observation, going back to Guarneri \cite{g} and further elaborated by several authors, is that quantum dynamical \textit{lower} bounds can be obtained from continuity properties of spectral measures (e.g., uniform H\"older continuity \cite{c,g}, non-singularity with respect to $x^\alpha$-Hausdorff measures \cite{l}, or non-singularity with respect to more general Hausdorff measures \cite{l2}).

The Jitomirskaya-Last extension \cite{jl1,jl2} of Gilbert-Pearson subordinacy theory provides an extremely useful tool for studying such continuity questions for one-dimensional Schr\"odinger operators (see also \cite{dkl,dl} for supplementary results in the whole-line case). It relates these questions to a study of generalized eigenfunctions and hence makes the subject quite accessible.

It is therefore not surprising that this method has been successfully applied to several models of physical relevance \cite{d1,dkl,dl,jl1,jl2}. Most of these results have dealt with models related to the Fibonacci Hamiltonian, the central object in the spectral theory of one-dimensional quasicrystals.

On the other hand, it is known that quantum dynamical \textit{upper} bounds do not follow from singularity properties of spectral measures alone (cf.\ \cite{djls,kl}). In fact, until very recently, no non-trivial quantum dynamical upper bounds were known for operators with purely singular continuous spectrum.

The article \cite{kkl} by Killip, Kiselev, and Last provides a general criterion for upper bounds in the context of one-dimensional Schr\"odinger operators. Their approach is based on \cite{jl1}, refines the Jitomirskaya-Last method, and relates upper bounds to growth properties of generalized eigenfuctions. These authors apply their method to the Fibonacci operator and hence establish anomalous transport behavior for the model: a property that had been expected and conjectured for a long time. 

In fact, the Fibonacci model consists of a family of operators and the dynamical result of \cite{kkl} holds for one member of this family. Our purpose here is to extend the result to the entire family. This is of physical importance since a quasicrystal is modelled by an LI-class -- a family of models with identical \textit{local} structure. It is expected that most relevant properties should hold for the entire family (or at least almost surely with respect to the unique ergodic measure that can usually be associated with such a family); see \cite{b} for further discussion of this issue.

Let us present the models we will be dealing with in this paper. We will study discrete one-dimensional Schr\"odinger operators

\begin{equation}\label{oper}
(H\phi) (n) = \phi(n+1) + \phi(n-1) + V(n) \phi(n)
\end{equation}
in $\ell^2(\Z)$ with Fibonacci potential

\begin{equation}\label{poten}
V(n) = \lambda v_\theta(n),
\end{equation}
where $\lambda > 0$ and $v_\theta$ is given by

\begin{equation}\label{vtheta}
v_\theta (n) = \chi_{[1-\omega , 1)} (n\omega + \theta \mod 1)
\end{equation}
with $\omega$ being the inverse of the golden mean, that is, $\omega = (\sqrt{5} - 1)/2$. It is well known, and can be shown using minimality of the hull and strong approximation, that the spectrum of $H$ is independent of the phase $\theta$, that is, for every $\lambda$ there is a set $\Sigma_\lambda \subseteq \R$ such that for every $\theta$, $\sigma(H) = \Sigma_\lambda$. We refer the reader to the survey articles \cite{d2,s2} for general information on the spectral theory of Schr\"odinger operators with Fibonacci and related potentials.

\section{Statement of the Main Result}

In this section we state our main result, Theorem~\ref{main}. Before doing so we introduce some notation.

Let $H$ be a discrete one-dimensional Schr\"odinger operator of the form \eqref{oper}. The unitary group $e^{-itH}$ associated with $H$ gives the time evolution of solutions to the time-dependent Schr\"odinger equation. Let $\delta_n$ be the element of $\ell^2(\Z)$ which is supported at $n \in \Z$ and obeys $\delta_n(n) = 1$. Given a function $\psi : \Z \rightarrow \C$ and $L > 0$, we define
$$
\| \psi \|_L^2 = \sum_{n = - \lfloor L \rfloor}^{\lfloor L \rfloor} | \psi (n) |^2 + (L - \lfloor L \rfloor ) \left( | \psi ( -\lfloor L \rfloor -1 ) |^2 + | \psi ( \lfloor L \rfloor + 1 ) |^2 \right).
$$
Finally, given a function $A : [0, \infty ) \rightarrow \R$, we define
$$
\langle A(t) \rangle_T = \frac{2}{T} \int_0^\infty e^{-2t/T} A(t) \, dt.
$$

Now let the potential $V$ of $H$ be given by \eqref{poten} and \eqref{vtheta}. Then, our main result reads as follows:

\begin{theorem}\label{main}
There are constants $C_1, C_2 ,G > 0$ such that for every $\lambda > 8$ and every $\theta$, we have

\begin{equation}\label{dynbound}
\left\langle \| e^{-itH} \delta_1 \|^2_{C_1 T^{p(\lambda)}} \right\rangle_T \ge G \mbox{ for every } T > 0,
\end{equation}
where $p(\lambda) = C_2 (\log \lambda)^{-1} (1 + O((\lambda \log \lambda)^{-1})$.
\end{theorem}

\noindent\textit{Remarks.} (a) The physical interpretation of \eqref{dynbound} is the following. At any time $T$, the averaged probability to find the particle in a ball of radius $C_1 T^{p(\lambda)}$ is uniformly bounded away from zero. This gives an upper bound on the \textit{slow} part of the wavepacket time evolution.\\[1mm]
(b) This theorem was shown by Killip et al.\ for the particular case $\theta = 0$ \cite{kkl}. The key ingredient in our extension to arbitrary phase $\theta$ will be a fine analysis of the subword structure of the sequences $v_\theta$, regarded as infinite words over the alphabet $\{0,1\}$. Once this subword structure is sufficiently well understood, we will be able to proceed along lines similar to the ones in the proof for the case $\theta = 0$.\\[1mm]
(c) As in \cite{kkl}, the theorem, while stated for initial vector $\delta_1$, can be recast for other initial vectors from $\ell^2(\Z)$.\\[1mm]
(d) Killip et al.\ also prove a quantum dynamical lower bound which, combined with the upper bound, shows that the asymptotic behavior for large coupling is optimal. That the lower bound extends to all phases is essentially contained in \cite{dkl} and hence will not be discussed here.

\medskip

The organization of this article is as follows. In the next section we discuss the combinatorics of the sequences $v_\theta$ and prove the key results that will enable us to pursue the strategy suggested by the general theorem on dynamical upper bounds from \cite{kkl}. We then give a proof of Theorem~\ref{main} in Section~4.

\section{Combinatorics of the Fibonacci Sequence}

In this section we analyze the local structure of the sequences $v_\theta$. We show in particular that the traces of the transfer matrices over intervals of length $F_k$, where $(F_k)_{k \in \N}$ is the sequence of Fibonacci numbers, exhibit a rather strong invariance property with respect to variation of the phase. This will then allow us to adapt the proof of Killip et al.\ to the general case.

Let us first recall some combinatorial notions. As general references we want to mention \cite{loth1,loth2}. Let $A$ be a finite set, called the alphabet, and denote by $A^*,A^\N,A^\Z$ the set of finite, one-sided infinite, and two-sided infinite words over $A$, respectively. If a word $w$ can be written as $w = w_1 w_2$ with words $w_1, w_2$, we call $w_1$ a prefix of $w$ and $w_2$ a suffix of $w$. Given any word $w$, we denote by $P_w$ the set of its finite subwords, and by $P_w(n)$ the set of its finite subwords of length $n$, $n \in \N$. Write $p_w(n)$ for the cardinality of $P_w(n)$; the function $p_w : \N \rightarrow \N$ is called the complexity function associated with $w$. If the word $w$ is infinite and uniformly recurrent (i.e., each of its finite subwords occurs infinitely often and with bounded gaps), we define the hull associated with $w$ by $\Omega_w = \{ s \in A^\Z : P_s = P_w \}$. 

In our concrete setting, the alphabet will be given by $A = \{0,1\}$ and the word $w$ will be given by the restriction of $v_0$ to $\N$, that is, $w = v_0(1) v_0(2) v_0(3) \ldots \in \{0,1\}^\N$. The word $w$ is called the Fibonacci sequence and its combinatorial properties have been studied extensively. For example, it is well known that its complexity function is given by

\begin{equation}\label{complex}
p_w (n) = n+1 \mbox{ for every } n.
\end{equation}
Moreover, it is also well known that $w$ is uniformly recurrent and that for every $\theta$, we have $v_\theta \in \Omega_w$. In particular, we have 

\begin{equation}\label{samewords}
P_{v_\theta} = P_w \mbox{ for every } \theta.
\end{equation}

Our goal is to study the sets $P_w(F_k)$, $k \in \N$, where $F_k$ is the $k$th Fibonacci number, that is, $F_{-1} = F_0 = 1$, $F_k = F_{k-1} + F_{k-2}$ for $k \ge 1$. One element of $P_w(F_k)$ is certainly given by the prefix $s_k$ of $w$ of length $F_k$. As is well known, the words $s_k$, $k \in \N$ obey recursive relations. Since this fact is crucial to our proof, we recall this result briefly. Consider on the alphabet $A = \{0,1\}$ the substitution $S : A \rightarrow A^*$ given by $S(0) = 1$, $S(1) = 10$. Extend this mapping morphically to $A^*$ and to $A^\N$. Then, we have (cf.\ \cite{s1})

\begin{equation}\label{alterdef}
s_k = S^k(1) \mbox{ for every } k \ge 0.
\end{equation}
Moreover, it follows from the substitution rule that

\begin{equation}\label{skrec}
s_k = s_{k-1} s_{k-2}.
\end{equation}

Our first observation is very simple:

\begin{lemma}\label{lastsymb}
For $k \ge 1$, the word $s_k$ has suffix $01$ if $k$ is even and it has suffix $10$ if $k$ is odd.
\end{lemma}

\begin{proof}
This follows immediately from $s_1 = 10$, $s_2 = 101$, and the recursion \eqref{skrec}.
\end{proof}

Define $\overline{\cdot} : A \rightarrow A$ by $\overline{0} = 1$, $\overline{1} = 0$. Our next goal is to list all elements of $P_w(F_k)$ explicitly. Write $s_k = s_k^{(1)} \ldots s_k^{(F_k)}$ with $s_k^{(i)} \in A$, $1 \le i \le F_k$.

\begin{lemma}\label{list}
For every $k \ge 0$, the $F_k + 1$ elements of $P_w(F_k)$ are given by

\begin{itemize}
\item The $F_k$ cyclic permutations of $s_k$ which are mutually distinct, and
\item the word $\overline{s_k^{(F_k)}} s_k^{(1)} \ldots s_k^{(F_k - 1)}$.
\end{itemize}
\end{lemma}

\begin{proof}
As remarked above, the word $w$ is the ``limit'' of the words $s_k$, in the sense that each $s_k$ is a prefix of $w$ and their lengths tend to infinity. Moreover, by \eqref{alterdef} we have the self-similarity property 

\begin{equation}\label{selfsim}
S(w) = w.
\end{equation}
The word $w$ begins with $10110 \ldots$ and hence we get from \eqref{alterdef} and \eqref{selfsim}  

\begin{equation}\label{kpart}
w = s_k s_{k-1} s_k s_k s_{k-1} \ldots \mbox{ for every } k \ge 0.
\end{equation}
Write $a$ for the rightmost symbol of $s_k$. Then, using Lemma~\ref{lastsymb}, we have that the word
$$
\begin{array}{ccc}
s_{k-1} & s_k & s_k \\
\fbox{\hspace{1cm} $| \, \overline{a}$} & \fbox{\hspace{1.6cm} $| \, a$} &\fbox{\hspace{1.6cm} $| \, a$} 
\end{array}
$$
belongs to $P_w$. In particular, all the words listed in the assertion of the lemma belong to $P_w(F_k)$. Finally, to conclude the proof all we have to show is that the $F_k$ cyclic permutations of $s_k$ are mutually distinct because by \eqref{complex} there are only $F_k + 1$ words in $P_w(F_k)$ and the list contains $F_k + 1$ words which are mutually distinct. To do so, let us first note that using the recursion $F_k = F_{k-1} + F_{k-2}$ one can prove by induction that

\begin{equation}\label{niceprop}
(-1)^{k-1} ( F_{k-2} F_k - F_{k-1}^2) = 1 \; \mbox{ for every } k \ge 1.
\end{equation}
Define, for $k \ge 0$, the height $h(s_k)$ of $s_k$ by $h(s_k) = $ number of $1$'s in $s_k$. It follows from the definition and \eqref{skrec} that $h(s_k) = F_{k-1}$. We can therefore interpret \eqref{niceprop} as 
$$
(-1)^{k-1}(h(s_{k-1}) |s_k| - h(s_k) |s_{k-1}|) = 1 \; \mbox{ for every } k \ge 1,
$$
from which we get that for $k \ge 0$, $|s_k|$ and $h(s_k)$ are relatively prime. This implies that the cyclic permutations of $s_k$ are mutually distinct, for otherwise $s_k$ could be written as a power of some shorter word, contradicting the above observation.
\end{proof}

We see that there is only one word $b_k$ in $P_w(F_k)$ which is not a cyclic permutation of $s_k$ and it can be described explicitly. In particular, it follows from Lemma~\ref{lastsymb} and Lemma~\ref{list} that the following holds:

\begin{equation}\label{firstsym}
\mbox{The leftmost symbol of $b_k$ is } \left\{ \begin{array}{ll} 0 & \mbox{if $k$ is even,} \\ 1 & \mbox{if $k$ is odd,} \end{array} \right. 
\end{equation}
and 

\begin{equation}\label{lastsym}
\mbox{The rightmost symbol of $b_k$ is } \left\{ \begin{array}{ll} 1 & \mbox{if $k$ is even,} \\ 0 & \mbox{if $k$ is odd.} \end{array} \right. 
\end{equation}

Denote by $s_k^\theta$ the word $v_\theta (1) v_\theta (2) \ldots v_\theta(F_k)$ and by $t_k^\theta$ the word $v_\theta (-F_k + 1) v_\theta (-F_k + 2) \ldots v_\theta(0)$. We can now state our main combinatorial result:

\begin{prop}\label{cyclic}
For every $\theta$, we have that $s_k^\theta$ is a cyclic permutation of $s_k$ for all $k$ odd or for all $k$ even. The same statement is true for $t_k^\theta$.
\end{prop}

\begin{proof}
Fix $\theta$. If $v_\theta(1) = 0$, then by \eqref{firstsym}, $s_k^\theta$ is conjugate to $s_k$ for all $k$ odd. Similarly, if $v_\theta(1) = 1$, then by \eqref{firstsym}, $s_k^\theta$ is conjugate to $s_k$ for all $k$ even. A completely analogous argument, using \eqref{lastsym}, yields the claim for $t_k^\theta$.
\end{proof}

\section{Proof of Theorem~\ref{main}}

In this section we give a proof of Theorem~\ref{main}. We will translate the combinatorial results of the previous section to statements on transfer matrix traces and then to statements on their norms. We can then employ a general theorem of \cite{kkl} which yields the claimed dynamical bound.

Let us describe the consequences of Proposition~\ref{cyclic} for the traces of certain transfer matrices. For $n \ge 1$, $E \in \R$, and $\lambda, \theta$ as above, let
$$
M(n,E,\lambda,\theta) = T(n,E,\lambda,\theta) \times \cdots \times T(1,E,\lambda,\theta),
$$
where for $m \in \Z$,
$$
T(m,E,\lambda,\theta) = \left( \begin{array}{cc} E - \lambda v_\theta(m) & -1 \\ 1 & 0 \end{array} \right).
$$
Let $x_k (E,\lambda,\theta) = \tr M(F_k,E,\lambda,\theta)$. 

\begin{prop}\label{traceequalities}
For every $\lambda,\theta$, we have that 

\begin{equation}\label{traceeq}
x_k(E,\lambda,\theta) = x_k(E,\lambda,0) \mbox{ for every } E \in \R
\end{equation}
holds for all $k$ odd or for all $k$ even. In particular,

\begin{equation}\label{deriveq}
\frac{\partial}{\partial E} x_k(E,\lambda,\theta) = \frac{\partial}{\partial E} x_k(E,\lambda,0) \mbox{ for every } E \in \R
\end{equation}
holds for all $k$ odd or for all $k$ even. 
\end{prop}

\begin{proof}
This is an immediate consequence of Proposition~\ref{cyclic} and the invariance of the trace of a product with respect to cyclic permutations of the factors.
\end{proof}

Similarly, for $n \le -1$, $E \in \R$, and $\lambda, \theta$ as above, let
$$
M(n,E,\lambda,\theta) = T(n+1,E,\lambda,\theta)^{-1} \times \cdots \times T(0,E,\lambda,\theta)^{-1}
$$
and define $y_k (E,\lambda,\theta) = \tr M(-F_k,E,\lambda,\theta)$. 

\begin{prop}\label{traceequalities2}
For every $\lambda,\theta$, we have that 

\begin{equation}\label{traceeq2}
y_k(E,\lambda,\theta) = x_k(E,\lambda,0) \mbox{ for every } E \in \R
\end{equation}
holds for all $k$ odd or for all $k$ even. In particular,

\begin{equation}\label{deriveq2}
\frac{\partial}{\partial E} y_k(E,\lambda,\theta) = \frac{\partial}{\partial E} x_k(E,\lambda,0) \mbox{ for every } E \in \R
\end{equation}
holds for all $k$ odd or for all $k$ even. 
\end{prop}

\begin{proof}
We have

\begin{eqnarray*}
y_k (E,\lambda,\theta) & = & \tr M(-F_k,E,\lambda,\theta)\\
& = & \tr \left( T(-F_k + 1,E,\lambda,\theta)^{-1} \times \cdots \times T(0,E,\lambda,\theta)^{-1} \right)\\
& = & \tr \left( T(0,E,\lambda,\theta) \times \cdots \times T(-F_k+1,E,\lambda,\theta) \right),
\end{eqnarray*}
where in the last step we have used that the determinant is one and hence the trace is invariant with respect to inverting the matrix. By Proposition~\ref{cyclic} and invariance of the trace with respect to cyclic permutations, for all even $k$ or for all odd $k$, the right-hand side is equal to $x_k(E,\lambda,0)$ for all $E$.
\end{proof}

We denote, for $L \ge 1$,
$$
\| M(E,\lambda,\theta) \|_L^2 = \sum_{n = 1}^{\lfloor L \rfloor} \| M(n,E,\lambda,\theta) (n) \|^2 + (L - \lfloor L \rfloor ) \left( \| M(\lfloor L \rfloor + 1, E, \lambda, \theta ) \|^2 \right)
$$
and, for $L < 1$, $\| M(E,\lambda,\theta) \|_L^2$ is defined analogously. Then the following holds:

\begin{prop}\label{normbound}
For every $\lambda > 8$, there are constants $C,\zeta > 0$ such that for every $\theta$, every $L$, and every $E \in \Sigma_\lambda$, we have
$$
\| M(E,\lambda,\theta) \|_L^2 \ge C |L|^\zeta.
$$
\end{prop}

\begin{proof}
Fix $\lambda > 8$ and some arbitrary $\theta$. Using a well-known formula from \cite{toda} (cf.~Section~6 of \cite{kkl}), one can show that

\begin{equation}\label{lowernorm}
4 \| M(E,\lambda,\theta) \|_{F_k}^3 \ge \left| \frac{\partial}{\partial E} x_k(E,\lambda,\theta) \right| \; \mbox{ for every } E.
\end{equation}
By Proposition~\ref{traceequalities}, for all even $k$ or for all odd $k$, the right-hand side is equal to $\frac{\partial}{\partial E} x_k(E,\lambda,0)$. We consider the case where this equality holds for all even $k$; the other case is similar. Proposition~5.2 of \cite{kkl} (which is derived from Raymond \cite{r}) says that we have 

\begin{equation}\label{lowerderiv}
\left| \frac{\partial}{\partial E} x_k(E,\lambda,0) \right| \ge \xi(\lambda)^{k/2} \; \mbox{ for every } E \in \Sigma_\lambda
\end{equation}
with $\xi(\lambda) = \lambda (1 + O(\lambda^{-1}))$ for $\lambda > 8$. Consider $L$ with $F_{2k} \le L < F_{2(k+1)}$. Since, for large $k$, $F_{2k} \sim \omega^{-2k}/\sqrt{5}$, it follows from \eqref{lowernorm} and \eqref{lowerderiv} that
$$
\| M(E,\lambda,\theta) \|_L^2 \ge \| M(E,\lambda,\theta) \|_{F_{2k}}^2 \ge C' |L|^\zeta \; \mbox{ with } \zeta = \frac{\log \xi(\lambda)}{3 \log (\omega^{-2})}.
$$
By adjusting the constant, we can cover the case of small positive $L$. Notice that both $C$ and $\zeta$ can be chosen uniformly in the phase $\theta$. A similar proof, using Proposition~\ref{traceequalities2}, works on the left half-line.
\end{proof}

\begin{proof}[Proof of Theorem~\ref{main}] 
Having established Proposition~\ref{normbound}, we can now proceed analogously to \cite{kkl}. Since from this point on the proof is very similar to what is done in \cite{kkl}, we refer the reader to Section~6 of that paper for further details. 
\end{proof}


\begin{thebibliography}{10}  

\bibitem{b} M.\ Baake, A guide to mathematical quasicrystals, in \textit{Quasicrystals}, Eds.~J.-B.~Suck, M.~Schreiber, and P.~H\"au{\ss}ler, Springer, Berlin (1999), in press; preprint (math-ph/9901014)
\bibitem{c} J.\ M.\ Combes, Connections between quantum dynamics and spectral properties of time-evolution operators, in \textit{Differential Equations with Applications to Mathematical Physics}, Eds.~W.~F.~Ames, E.~M.~Harrell, and J.~V.~Herod, Academic Press, Boston (1993), 59--68
\bibitem{d1} D. Damanik, $\alpha$-continuity properties of one-dimensional quasicrystals, \textit{Commun.\ Math.\ Phys.} {\bf 192} (1998), 169--182
\bibitem{d2} D.\ Damanik, Gordon-type arguments in the spectral theory of one-dimensional quasicrystals, in \textit{Directions in Mathematical Quasicrystals}, Eds.~M.~Baake and R.~V.~Moody, CRM Monograph Series {\bf 13}, AMS, Providence, RI (2000), 277--305
\bibitem{dkl} D.\ Damanik, R.\ Killip, and D.\ Lenz, Uniform spectral properties of one-dimensional quasicrystals, III. $\alpha$-continuity, \textit{Commun.\ Math.\ Phys.} {\bf 212} (2000), 191--204
\bibitem{dl} D.\ Damanik and M.\ Landrigan, Log-dimensional spectral properties of one-dimensional quasicrystals, preprint (2001)
\bibitem{djls} R.\ del Rio, S.\ Jitomirskaya, Y.\ Last, and B.\ Simon, Operators with singular continuous spectrum. IV. Hausdorff dimensions, rank one perturbations, and localization, \textit{J.\ Anal.\ Math.} {\bf 69} (1996), 153--200
\bibitem{gp} D.\ J.\ Gilbert and D.\ B.\ Pearson, On subordinacy and analysis of the spectrum of one-dimensional Schr\"odinger operators, \textit{J.\ Math.\ Anal.\ Appl.} {\bf 128} (1987), 30--56
\bibitem{g} I.\ Guarneri, Spectral properties of quantum diffusion on discrete lattices, \textit{Europhys.\ Lett.} {\bf 10} (1989), 95--100
\bibitem{jl1} S.\ Jitomirskaya and Y.\ Last, Power-law subordinacy and singular spectra. I. Half-line operators, \textit{Acta Math.} {\bf 183} (1999), 171--189
\bibitem{jl2} S.\ Jitomirskaya and Y.\ Last, Power-law subordinacy and singular spectra. II. Line operators, \textit{Commun.\ Math.\ Phys.} {\bf 211} (2000), 643--658
\bibitem{kkl} R.\ Killip, A.\ Kiselev, and Y.\ Last, Dynamical upper bounds on wavepacket spreading, preprint (2001), available from mp-arc (01--457)
\bibitem{kl} A.\ Kiselev and Y.\ Last, Solutions, spectrum, and dynamics for Schr\"odinger operators on infinite domains, \textit{Duke Math.\ J.} {\bf 102} (2000), 125--150
\bibitem{l2} M.\ Landrigan, Log--dimensional properties of spectral measures, Ph.~D.~thesis, UC~Irvine (2001)
\bibitem{l} Y. Last, Quantum dynamics and decompositions of singular continuous spectra, \textit{J.\ Funct.\ Anal.} {\bf 142} (1996), 406--445
\bibitem{loth1} M.\ Lothaire, \textit{Combinatorics on words}, Cambridge University Press, Cambridge (1997)
\bibitem{loth2} M.\ Lothaire, \textit{Algebraic combinatorics on words}, in preparation
\bibitem{r} L.\ Raymond, A constructive gap labelling for the discrete Schr\"odinger operator on a quasiperiodic chain, preprint (1997)
\bibitem{s1} A.\ S\"ut\H{o}, The spectrum of a quasiperiodic Schr\"odinger operator, \textit{Commun.\ Math.\ Phys.} {\bf 111} (1987), 409--415
\bibitem{s2} A.\ S\"ut\H{o}, Schr\"odinger difference equation with deterministic ergodic potentials, in \textit{Beyond Quasicrystals} (Les Houches, 1994), Eds.~F.~Axel and D.~Gratias, Springer, Berlin (1995), 481--549
\bibitem{toda} M.\ Toda, \textit{Theory of Nonlinear Lattices}, Springer, Berlin (1989)


\end{thebibliography}
\end{document}